# THE EXPLORATORY RESEARCH OF THE EFFECT COMMUNICATION MODEL AND EFFECT IMPROVING STRATEGY OF INTERACTIVE ADVERTISING

Ding Wanxing1

[1]College of Communication and Art Design, University of Shanghai for Science and Technology, Shanghai, China
270258632@qq.com

## ABSTRACT

*Interactive advertising which characterized by interactivity has become the mainstream of advertising by gradually replacing traditional one-way advertising during the new media era. This paper has obeyed the research outline below, literature review, background description, assumption proposed, and empirical analysis.*

*Therefore, this paper proposed the communication model of interactive advertising and drawn two related important conclusions, interactivity has brought positive effect to advertising communication, different type of consumers tend to use different interactive options in different ways. Furthermore, this paper also presented three related optimization strategies to improving the communication of interactive advertising, namely, 1.changing communication model from one-way to two-way, 2.renovating new communication process and effect-generated path, 3.renovating new strategy portfolio to improving the communication effect of interactive advertising.*

## 1. INTRODUCTION

Interactive advertising which characterized by interactivity has become the mainstream of advertising by gradually replacing traditional one-way advertising during the new media era. This paper has obeyed the research outline below, literature review, background description, assumption proposed, and empirical analysis. Specifically, this paper has two aspects of research objectives and research level, first of all, whether interactivity is the key factor to improving the advertising effect or not, second, whether the consumer characteristics effected interaction options' usage or not. Thus, this paper has used the method of empirical research with the case of New Balance, and aimed to exploring the internal communication model of interactive advertising through the mathematical analysis.

## 2. LITERATURE REVIEW

### 2.1. Interactive Advertising

Broadly speaking, interactive advertising (IA) communication includes advertisement on many kinds of platforms and channels such as Internet, CD-ROMs and the kiosks around stations. In this level, interactive advertising includes any ads which audience were required to make reply, even banner ads belong to interactive ads under this context because it required to click.

However, the research topic of this paper focus on the Internet interactive advertising, especially the household FMCG brands interactive advertising, for example, New Balance running shoes.

There were mainly three reasons for choosing FMCG brands, first of all, minimize the research scope to maximize the research representative, secondly, different kinds of ads such as commercials and PSAs bearing different advertising appeals, designs as well as production; finally, the experimental data sample of this paper were collected from college students, taking students purchasing intension and ability into consideration, chosen FMCG brands interactive advertising as the research target ensured the higher scientific and representative statistical analysis compared to high-grade durable goods such as real estate, machinery manufacturing.

**2.2. Communication Model**

Model, a simplified theory form of representation of reality, which has four main functions, first of all, organizational function, namely to reveal the order and the interrelationship among different systems, secondly, interpretational function, which means providing information in a simply and clear way while in other ways these information would become complex or ambiguous, thirdly, inspirational function, namely to guide people to focus on the core aspects of a process or a system, and finally, prediction function, which means to make prediction of the process and result through a model.

Starting from Lasswell's 5W liner model, the research of communication model has made the first breakthrough by introduce the concept of structure, process into the research area of communication in details. Later, Schramm, Dans both referred to communication was not the one-way process from sender to receiver, but the two-way and recycling process by introducing receiver feedback mechanism, therefore, cybemetic model of communication was known as emphasizing receiver's initiatives. Until Riley couple introduce social structure as external factor into communication model theory, the advent of social system model has supplemented and completed the communication model theory.

Thus, the three main communication models above were not only revealed the significance to research the interactive advertising communication model, but also suggested some research tips of communication model in this paper required to pay attention to.

**2.3 Communication Effect**

Communication effect refers to 'after reaching receiver via certain media channel, message which sent by sender will influences receiver's thinking and behaviour'. Macraw has made further comments on communication effect, which referred to 'communication effect can be divided into micro effect and macro effect, changing effect and stable effect, cumulative effect and non-cumulative effect, short-term effect and long-term effect, attitude effect, cognition effect and behaviour effect'.

Specifically, on this study, from the perspective of micro effect of individual audience, the author aimed to explain the media impact of a certain particular class of social system at the macro level by collecting some relative homologous sample quantity; at the same time, the design of this experiment emphasize the non-cumulative and short-term effect, which means this paper analyse the communication effect in an immediate and relatively short-term. Furthermore, from the aspect of effect's content, this paper not only included the changing and stable effect, but also contained attitude effect, cognition effect and behaviour effect. This kind of subdivision on effects' research was also one of the innovations of this paper.

**2.4 Consumer Characteristics**

Consumer can be categorized from different consumer characteristics, namely the level of consumer need for cognition, need for emotion and product involvement. These three standards to describe consumer characteristics were used as the independent variables of this paper.

Need for cognition is designed to measure an individual's tendency to engage in and enjoy effortful cognitive endeavours. In this paper, need for cognition was measured with the 18-item, short-form version developed by Cacioppo, petty and Kao, and respondents were asked to state their agreement or disagreement with these 18 statements on a five-point scale.

Need for emotion referred to tendencies to process affective emotional stimuli, also, its' definition was the tendency or propensity for individuals to seek out emotional situations, enjoy emotional stimuli and exhibit a preference to use emotion in interacting with the world. In this paper, need for emotion was measured by seven-item short-form version and respondents stated their agreement or disagreement on a five-point scale.

Product involvement referred as respondents' overall evaluation of how important the product is to their life. In this paper, the author took Zaichkowsky's set of 10 seven-point, semantic differential statements as well as end-points describing product involvement (important/unimportant, relevant/irrelevant etc.).

## 3. MAIN RESEARCH

### 3.1 Research Objectives

Specifically, there are two research objectives in this paper, whether the interactive characteristics will affect the communication effect of advertising; whether the characteristics of consumers will bring differences on the usage of interactive advertising.

Based on these two research objectives, this paper has certain values in academic theory level as well as the actual operation level. In the level of academic theory, this paper aimed to provide a clear effect-generate path for the communication patterns of the interactive advertising, in order to integrate the research of advertising communication effect, also to enrich the research of interactive advertising.

On the other hand, in the level of actual operation of interactive advertising, the empirical research in this paper not only explored the communication and sales results after the interaction, but also integrated the entire research skeleton from consumer to product sales, which provided a whole complete research cycle concluded interactive advertising, consumer characteristics as well as the design of interactive dimensions and options.

### 3.2 Research Methods

For the purpose of this paper, the author advocated a scientific research routine of the integration of the traditional four science course, namely, normativism courses, positivism courses, pragmatism courses, and empiricism courses.

Specifically, on the one hand, the author used normative and positivism course, namely using experimental investigation to analyse mathematically and empirically, inducing the internal mechanism and communication model of interactive advertising, which brought experiences to scientific theory. On the other hand, this paper adopted pragmatism courses, which refers to putting forward countermeasures and suggestions to solve problems, that is to say, proposing optimization strategy to improving the communication effect of interactive advertising.

This paper used comparative experimental investigation method. In specific, the author divided research subjects into two groups, experimental group and control group. First of all, taking the students' number of University of Shanghai for Science and Technology as sampling frame and sampled 90 students as research subjects according to the table of random numbers; secondly, arrange research groups, sorting by those samples surname stroke, then the singular arranging to experimental group, while the dual arranging to control group, and they would be required to browse interactive advertising website with high-interactivity and with low-interactivity respectively, and after browsed respondents were asked to complete investigation questionnaire.

Furthermore, in order to control variables to analyse the communication model accurately, the author chose the same brand (New Balance), the same style (VAZEE Collection Running Shoes) and the same advertising content, which means, the comparative interactive advertising website with the same content but the different degree of interactivity, in order to exploring the interactivity's effect to the communication of interactive advertising. Therefore, the low-interactivity ads were simply a copy of the brands information with no added interactivity features, while the high-interactivity ads contained more types of interactivity features such as hypertext, comments form, linked visuals and so on.

Specially, the experimental investigation of this paper was field experiment, namely, experimental respondents could finish this experiment at anywhere they feel convenient and comfortable and in anytime they were free, which only appliance they need was a computer with Internet access. Thus, this field experiment highly restored the real online-shopping experience, there were no time limit as well as location limit of experiment, when respondents browsed the experimental website materials, respondents can enjoy browsing as long as they wish, while respondents finished the step of browsing website, they were asked to closing the website and could not open it again when they answering investigation questionnaire.

### 3.3 Background Analysis and Research Hypothesis

The relationship between interactivity and advertising effect was obvious and intricate through analysis of quantitative successful interactive advertising cases in our daily lives. First of all, from the perspective of the relationship of consumer and interaction, consumer characteristics were tightly related to using interactive options, for instance, social butterflies preferred involved in human-human interaction. Secondly, from the perspective of interactive dimensions, almost interactive advertising contained the inner interactions (mainly human-human interaction, human-information interaction) and the outer interactions that linked to outer websites. Thirdly, form the perspective of advertising effect, almost all of them were achieved broader participation, higher advertising comprehension, attitude towards advertising, brand, the website as well as the higher sales values and marketing values compared to traditional advertisings.

Further, interactive advertising effect and its' marketing values reflected around interactivity which is the core new media mechanism. On the one hand, higher degree of interactivity, which means much more possibility to reach sum non-zero effect, namely, interactivity was the key factor conveying various contact between different levels and different types, which take great advantage to formatting a multi-dimensional network that can be rely on to all kinds of advertising involvers, such as, advertisers and consumers, consumer and consumers could trust each other, and depended on each other. Such kind of multi-dimensional network would benefit to build a self-enhanced, mutual benefit relationship between consumers and advertisers. Take UNIQLO's Lucky line as one typical interactive advertising, the online queued interactive advertising appealed amounts of participants due to the discount coupons awards offline. In this case, win-win marketing has achieved, advertiser got the eyeballs and profits, while consumer got the discount, moreover, this kind of interaction urging consumer wants " more three wishes", namely, interactive more, to get more discount coupons, such kind of recursive self-improvement cycle promote the advertising effect under the core mechanism of interactivity. On the other hand, the more interaction, the better effect of advertising. Like Kevin Kelly pointed out in his book 'Mechanism', which referred to cumulative social organizations show something as the taste of pure mathematics which beyond good neighbor relations. This incremental system is based on information flow, which strains trust and competition into an inter-dependent networks, with the increase of these links, this enhanced and accelerated strength also increases accordingly. In interactive advertising, the incremental system of interactive effect is also based on the information flow which is given by interactive options design. The more interactive links, options and buttons, the more extensive consumer interaction degree will be, and the strength promoted by this effect increased correspondingly.

In summary, the author proposed the following six hypotheses.

H1.Interactivity can improve the advertising comprehension, namely, the advertising comprehension of high-interactivity advertising will be higher than the level of comprehension of the low-interactivity advertising.

H2. Interactivity can improve the advertising attitude, namely, the attitude toward advertising of high-interactivity advertising will be higher than the level of attitude of the low-interactivity advertising.

H3.Interactivity can improve the purchase intention, namely, the purchase intention of high-interactivity advertising will be higher than the level of purchase intention of the low-interactivity advertising.

H4.Consumer with High-Need for Cognition (HNFC) will use human-information interaction more than consumer with Low-Need for Cognition (LNFC).

H5.Consumer with High-Need for Emotion (HNFE) will use human-human interaction more than consumer with Low-Need for Emotion (LNFE).

H6.Consumer with High-Product Involvement (HPI) will use human-information interaction more than consumer with Low- Product Involvement (LPI).

Wherein, H1, H2, H3 were designed to exploring whether interactivity was the key factor to improving the advertising effect in terms of comprehension, attitude and purchase intention, whereas H4, H5, H6 aimed to research relationship between consumer and interaction usage, which combined subjective and objective aspects together, bring a comprehensive interactive advertising communication effect model form the former consumer to mediate interactive options as well as to latter communication effect.

## 4. EXPERIMENTAL STATISTICS ANALYSIS

### 4.1 Respondent Characteristics Analysis

The research used a convenience sample of 90 students form University of Shanghai for Science and Technology. Sixty percentage (n=54) were male and forty percentage (n=45) were female. Their ages ranged between 22 and 32 years old with the majority being between 24 and 25 years of age (n=73). Their majors in school varied from communication to engineering, automatic and so forth. The average time they stayed in school over two years. Therefore, the sample is more or less the typical student representatives.

### 4.2 Experimental Statistics Analysis

H1.Interactivity can improve the advertising comprehension, namely, the advertising comprehension of high-interactivity advertising will be higher than the level of comprehension of the low-interactivity advertising.

Comprehension was measured by six semantic judgement and two open questions, and about the semantic judgement questions, answer right scored one point, answer false scored zero point, the results are listed in tables 1, 2 and 3.

Tabel1. Summary of open questions

| N=90 | Q1. information of the browsed web site | Q2 feelings and suggestions for the browsed the web site |
|---|---|---|
| **H-interactivity（n=45）** | NB Running shoes n=45 | The web page is very cool and with good interaction n=21 |
| | NB VAZEE collections running shoes n=23 | Hope for more styles of shoes n=10 |
| | A total of three styles for NB running shoes n=37 | The website information clearly illustrated n=9 |
| | NB VAZEE evolutionary lightweight speed running shoes n=8 | Like NB much more after browse the web site n=7 |
| **L-interactivity (n=45)** | NB Running shoe n=45 | Page drag too long with badly interaction n=23 |
| | NB VAZEE collections running n=9 | Hope for seeing other consumers' evaluation n=13 |
| | A total of three styles for NB running shoes n=20 | The web site has good tone and style n=9 |
| | NB VAZEE evolutionary lightweight speed running shoes n=1 | The information on the web site is structured and completely n=7 |

Table2. Group statistics for the comprehension of high & low interactivity advertising

| | | N | Means | Standard deviation | Standard error of the mean |
|---|---|---|---|---|---|
| Comprehension | H-interactivity | 45 | 5.16 | .852 | .127 |
| | L-interactivity | 45 | 2.40 | .963 | .144 |

Tabel3. Independent Samples Test for the comprehension of high &low interactivity advertising

| | | Levene's test for equality of variances | | T test for Equality of Means | | | | | | |
|---|---|---|---|---|---|---|---|---|---|---|
| | | F | Sig. | t | df | Sig. | Mean Difference | Std. Error Difference | 95% Confidence Interval of the Difference | |
| | | | | | | | | | Lower | Upper |
| comprehension | Equal Variances assumed | .314 | .047 | 14.379 | 88 | .000 | 2.756 | .192 | 2.375 | 3.136 |
| | Equal Variances not assumed | | | 14.379 | 86.704 | .000 | 2.756 | .192 | 2.375 | 3.136 |

Base on the T-test results presented in Tables 3, the mean scores of comprehension level of low-interactivity group was 2.40, and the mean scores of comprehension level of high-interactivity group was obviously higher than the low-interactivity one. And the Levene's test for equality of variances showed statistically significant at the .05 level (F=0.314, p<0.05), so variance uneven of two samples, T=14.397, df=86.704., p<0.05. Therefore, there were statistically significant differences of comprehension levels caused by the interactivity degree, hypothesis 1 was supported.

Additionally, interactivity indeed influenced respondents' comprehension to advertisement message in different dimensions through an important way. Firstly, respondents' comprehension to ads was enhanced inevitably through continuous interactions. Second of all, the level and the depth of respondents' comprehension would be different caused by different level of website's interactivity, shown as table2, Although respondents from two groups could understand the research brand, namely, New Balance Running Shoes, further and deep comprehension to ads showed significant differences, specifically, there are 8 respondents recalled the "NB evolutionary lightweight running shoes" which accounting for 18 percentage of high-interactivity group, while only 1 respondent could recall this message which just accounting for 2 percentage of low-interactivity group. Thereby, 23 respondents from high-interactivity group recalled "VAZEE running shoes from NB" accurately, while only 9 respondents from low-interactivity group could recalled the same message, 37 respondents from high-interactivity group recalled three different types of NB VAZEE running shoes which accounting for 82 percent of the group, while only 20 respondents from low-interactivity group could recalled the same message, the comprehension between the two groups were significant which close to 2 times. Last but not least, there were still some respondents expressed more appreciation and praise over the low-interactivity version. From their point of view, the low-interactivity version holds the reputation of clear website style and tone of design, complete information and the distinguished level system. The author convinced the phenomenon was associated with the respondents' characteristics like website experience closely, wherein, about 81 percentage (n=13) respondents scored lower than average in Need for Cognition (M=53.12), and approximately 94 percentage (n=15) respondents has lesser website experiences (M=5yrs).

H2. Interactivity can improve the advertising attitude, namely, the attitude toward advertising of high-interactivity advertising will be higher than the level of attitude of the low-interactivity advertising.

Table4. Group statistics for attitude evaluation of high & low interactivity advertising

|  |  | N | Means | Standard deviation | Standard error of the mean |
|---|---|---|---|---|---|
| Advertising attitude | H-interactivity | 45 | 24.27 | 5.833 | .869 |
|  | L-interactivity | 45 | 20.87 | 4.409 | .657 |

Table5. Independent Samples Test for the attitude of high & low interactivity advertising

|  |  | Levene's test for equality of variances | | T test for Equality of Means | | | | | | |
|---|---|---|---|---|---|---|---|---|---|---|
|  |  | F | Sig. | t | df | Sig. | Mean Difference | Std. Error Difference | 95% Confidence Interval of the Difference | |
|  |  |  |  |  |  |  |  |  | Lower | Upper |
| Attitude | Equal Variances assume | 5.698 | .019 | 3.120 | 88 | .002 | 3.400 | 1.090 | 1.234 | 5.566 |

| | | | | 3.120 | 81.905 | .003 | 3.400 | 1.090 | 1.232 | 5.568 |
|---|---|---|---|---|---|---|---|---|---|---|
| | Equal Variances not assumed | | | | | | | | | |

Based on the Tables 4 and 5, the mean scores of attitude evaluation level of low-interactivity group was 24.27, which was much higher than the mean scores of comprehension level of high-interactivity group (M=20.87). And the Levene's test for equality of variances showed statistically significant at the .05 level (F=5.595, p<0.05), so variance uneven of two samples, T=3.120, df=81.905., p<0.05. Therefore, there were statistically significant differences of attitude evaluation levels caused by the interactivity degree, hypothesis 2 was supported.

H3. Interactivity can improve the purchase intention, namely, the purchase intention of high-interactivity advertising will be higher than the level of purchase intention of the low-interactivity advertising.

Table6. Group statistics for purchase intension of high & low interactivity advertising

| | | N | Means | Standard deviation | Standard error of the mean |
|---|---|---|---|---|---|
| Purchase Intention | H-interactivity | 45 | 12.09 | 1.663 | .248 |
| | L-interactivity | 45 | 9.27 | 1.498 | .223 |

Table7. Independent Sample Test for purchase intension of high & low interactivity advertising

| | | Levene's test for equality of variances | | T test for Equality of Means | | | | | | |
|---|---|---|---|---|---|---|---|---|---|---|
| | | F | Sig. | t | df | Sig. | Mean Difference | Std. Error Difference | 95% Confidence Interval of the Difference | |
| | | | | | | | | | Lower | Upper |
| Purchase intention | Equal Variances assume | .006 | .040 | 8.458 | 88 | .000 | 2.822 | .334 | 2.159 | 3.485 |
| | Equal Variances not assumed | | | 8.458 | 87.065 | .000 | 2.822 | .334 | 2.159 | 3.485 |

Based on the Tables 6 and 7, the mean scores of purchase intention of low-interactivity group was 12.09, which was much higher than the mean scores of purchase intention of high-interactivity group (M=9.27). And the Levene's test for equality of variances showed statistically significant at the .05 level (F=0.006, p<0.05), so variance uneven of two samples, T=8.458, df=87.065, p<0.05. Therefore, there were statistically significant differences of purchase intention caused by the interactivity degree, hypothesis 3 was supported.

In sum up, interactivity indeed brought a significant improving effect to advertising effect as we predicted before, and in specific, interactivity is persuasive to helping improving comprehension, attitude evaluation and purchase intention gradually.

Furthermore, in addition to the impact of interactivity, whether consumer characteristics influenced the comprehension level of advertising. This part of research has vital significance to modify the practice of interactive advertising, namely, advertisers have to customize different interaction features to different consumers with different type of characteristics.

To research H4, H5 and H6, the author used a combination of cross 6*2 of related variables which listed in Tables 8.

Table 8. Combination of cross 6*2 of variables.

|  | **Human-Information interact** | **Human-Human interact** |
|---|---|---|
| **HNFC** | HNFC-Human-information interact | HNFC-Human-Human interact |
| **LNFC** | LNFC-Human-information interact | LNFC-Human-Human interact |
| **HNFE** | HNFE-Human-information interact | HNFE-Human-Human interact |
| **LNFE** | LNFE-information interact | LNFE-Human-Human interact |
| **HPI** | HPI-information interact | HPI-Human-Human interact |
| **LPI** | LPI-information interact | LPI-Human-Human interact |

Table 9. Group statistics (Need for cognition)

|  |  | N | Means | Standard deviation | Standard error of the mean |
|---|---|---|---|---|---|
| Human-Information interaction | HNFC | 45 | 19.33 | 3.431 | .511 |
|  | LNFC | 45 | 16.56 | 2.473 | .369 |
| Human-Human interaction | HNFC | 45 | 16.42 | 3.101 | .462 |
|  | LNFC | 45 | 17.89 | 3.231 | .482 |

Table 10. Independent Samples Test (Need for cognition)

|  |  | Levene's test for equality of variances | | T test for Equality of Means | | | | | | |
|---|---|---|---|---|---|---|---|---|---|---|
|  |  | F | Sig. | t | df | Sig. | Mean Difference | Std. Error Difference | 95% Confidence Interval of the Difference | |
|  |  |  |  |  |  |  |  |  | Lower | Upper |
| Human-information interaction | Equal Variances assume | 4.310 | .041 | 5.992 | 88 | .000 | 3.778 | .631 | 2.525 | 5.031 |
|  | Equal Variances not assumed |  |  | 5.992 | 80.001 | .000 | 3.778 | .631 | 2.523 | 5.033 |
| Human-human interaction | Equal Variances assume | .079 | .780 | -3.096 | 88 | .003 | -2.067 | .668 | -3.393 | -.740 |
|  | Equal Variances not assumed |  |  | -3.096 | 87.852 | .003 | -2.067 | .668 | -3.393 | -.740 |

Table 11. Group statistics (Need for emotion)

|  |  | N | Means | Standard deviation | Standard error of the mean |
|---|---|---|---|---|---|
| Human-Information interact | HNFE | 45 | 16.02 | 3.621 | .540 |
|  | LNFE | 45 | 17.07 | 3.422 | .510 |
| Human-Human interact | HNFE | 45 | 18.18 | 3.645 | .543 |
|  | LNFE | 45 | 16.73 | 3.063 | .457 |

Table 12. Group statistics (Product involvement)

|  |  | N | Means | Standard deviation | Standard error of the mean |
|---|---|---|---|---|---|
| Human-Information interact | HPI | 45 | 17.56 | 3.609 | .538 |
|  | LPI | 45 | 17.46 | 3.262 | .486 |
| Human-Human interact | HPI | 45 | 18.08 | 3.040 | .453 |
|  | LPI | 45 | 16.53 | 3.571 | .532 |

Table 13. Independent Samples Test (Need for emotion)

|  |  | Levene's test for equality of variances | | T test for Equality of Means | | | | | | |
|---|---|---|---|---|---|---|---|---|---|---|
|  |  | F | Sig. | t | df | Sig. | Mean Difference | Std. Error Difference | 95% Confidence Interval of the Difference | |
|  |  |  |  |  |  |  |  |  | Lower | Upper |
| Human-information Interaction | Equal Variances assume | .163 | .687 | -1.137 | 88 | .259 | -.844 | .743 | -2.320 | .632 |
|  | Equal Variances not assumed |  |  | -1.137 | 87.719 | .259 | -.844 | .743 | -2.320 | .632 |
| Human-Human interaction | Equal Variances assume | 4.428 | .038 | 2.035 | 88 | .045 | 1.444 | .710 | .034 | 2.855 |
|  | Equal Variances not assumed |  |  | 2.035 | 85.465 | .045 | 1.444 | .710 | .033 | 2.855 |

Table 14. Independent Samples Test (Product involvement)

| | | Levene's test for equality of variances | | T test for Equality of Means | | | | | 95% Confidence Interval of the Difference | |
|---|---|---|---|---|---|---|---|---|---|---|
| | | F | Sig. | t | df | Sig. | Mean Difference | Std. Error Difference | Lower | Upper |
| Human-information interaction | Equal Variances assume | 1.834 | .179 | -.276 | 88 | .783 | -.200 | .725 | -1.641 | 1.241 |
| | Equal Variances not assumed | | | -.276 | 87.118 | .783 | -.200 | .725 | -1.641 | 1.241 |
| Human-human interaction | Equal Variances assume | .505 | .479 | 2.638 | 88 | .010 | 1.844 | .699 | .455 | 3.234 |
| | Equal Variances not assumed | | | 2.638 | 85.810 | .010 | 1.844 | .699 | .455 | 3.234 |

Based on tables 9 to 14, we can infer consumer usage preference. From the level of the usage of human-information interaction, respondents with high-need for cognition (M=19.33)> respondents with high-product involvement (M=17.56) > respondents with low-product involvement (M=17.46) > respondents with low-need for emotion (M=17.07) > respondents with low-need for cognition (M=16.56) > respondents with high-need for emotion (M=16.02). From the level of the usage of human-human interaction, respondents with high-need for emotion (M=18.18) > respondents with high-product involvement (M=18.08) > respondents with low-need for cognition (M=17.89) > respondents with low-need for emotion (M=16.73) > respondents with low-product involvement (M=16.53) > respondents with high-need for cognition (M=16.42).

Meanwhile, from statistics of the mean distribution level showed in tables above, the following three conclusions can be drawn roughly. First of all, the usage level of human-information interaction was highest for respondents with high need for cognition, while the usage level of human-human interaction was highest for respondents with high need for emotion. Secondly, there was statistically significance between consumer with high or low product involvement in the level of the usage of human-human interaction rather than human-information interaction. Thirdly, from the perspective of the usage level of human-human interaction and human-human interaction, high-need for cognition always accompanying with the characteristics of low-need for emotion, high-need for emotion always accompanying with the characteristics of low-need for cognition. Those two characteristics above always occupy the two extremes of the scale, as shown below in figure1.

Figure1. Consumer characteristics scale

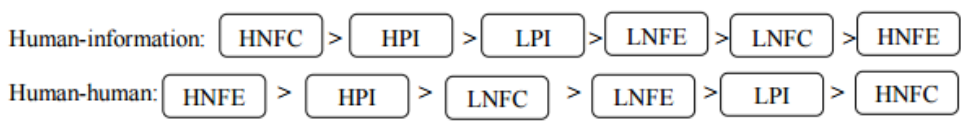

Furthermore, based on the T-test results showed in tables above, the mean usage level of human-information interaction of respondents with high-need for cognition (M=19.33) was higher than respondents with low-need for cognition (M=16.56), F=4.310, p<0.05, so variance uneven of two samples, then T=5.992, df=80.001, p<0.05. Therefore, there were statistically significant differences of human-information interaction usage to different consumers with high or low level of need for cognition, hypothesis 4 was supported.

And from the perspective of the usage of human-human interaction, the mean usage level of human-human interaction of respondents with high-need for emotion (M=18.18) was higher than respondents with low-need for emotion (M=16.73). F=4.428, p<0.05, so variance uneven of two samples, then T=2.035, df=85.465, p<0.05. Therefore, there were statistically significant differences of human-information interaction usage to different consumers with high or low level of need for emotion, hypothesis 5 was supported.

From the perspective of the usage of human-information interaction, the mean usage level of human-information interaction of respondents with high-product involvement (M=17.56) was higher than respondents with low- product involvement (M=17.46).But the t-test showed p>0.05, there was no statistically significant differences of human-information interaction usage to different consumers with high or low level of product involvement. Conversely, from the perspective of human-human interaction, the mean usage level of human-human interaction of respondents with high-product involvement (M=18.08) was higher than respondents with low product involvement (M=16.53). F=0.505, p<0.05, then refine T=2.638, df=88, p<0.05. Therefore, there were statistically significant differences of human-human interaction usage to different consumers with high or low level of product involvement. Hypothesis 6 was not supported.

In a word, hypotheses 4 and 5 was supported, consumer with High-Need for Cognition (HNFC) will use human-information interaction more than consumer with Low-Need for Cognition (LNFC). Consumer with High-Need for Emotion (HNFE) will use human-human interaction more than consumer with Low-Need for Emotion (LNFE). Hypothesis 6 was not supported, respondents with high-product involvement prefer use human-human interaction rather than human-information interaction.

## 5. CONCLUSIONS

### 5.1 Communication Model of Interactive Advertising

Overall, the author summarized the communication model of interactive advertising as well as the effect improving strategies showed in figure2.

Figure2. Communication model of interactive advertising

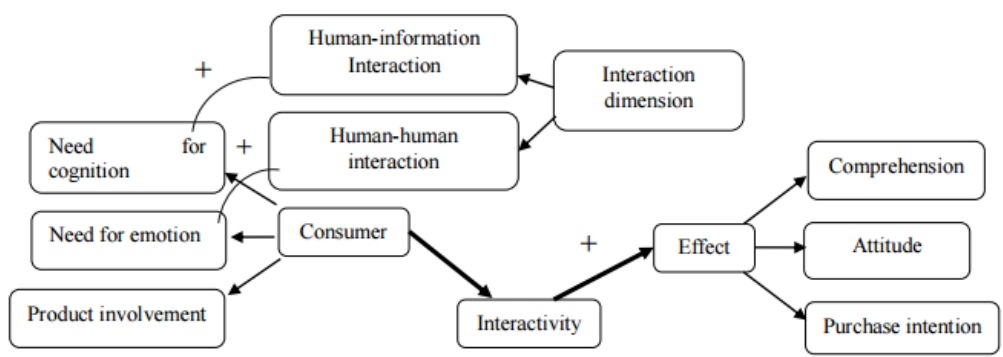

## 5.2 Effect Improving Strategy

5.2.1 Qualitative change of communication model: from 'one way' to 'two way', form 'be-persuaded' to 'self-persuaded'

The core mechanism of two-way communication is achieving the self-persuasion of communication and self enhancement of advertising effect. Just as Hovland pointed out in the research of persuasive communication: the effect of two sides persuasion (Both positive and negative information) is much better than single side persuasion (Only to say some information which will benefit themselves), Because once exposed to negative information, receiver effected by single side will easily defect, While receiver effected by two sides is much more firm. Which showed that the two sides persuasion have the vaccination effect , enhancing the immunity of receiver against the opposite opinion .It's the same to apply it to the advertising communication, when the advertising is no longer straightforward declare some one side information of product such as good quality, high value and elegant style, but also through the prizes or incentives of discount to encourage the audience's independent participation in interaction, then audience would understand the product better with diversified information during the process of interaction(including comments from other users, contact more information, and even bad comments). In that cases, audience felt was no more than be persuaded by advertisers, but be persuaded by themselves on the basis of understanding, at the same time, with the influence of vaccination effect, the audience have strong ability to resist to negative information, such kind of persuasion effect tended to be more strong and long-lasting, the accompanying brand viscosity also became more stronger too, which brought the qualitative change of communication model from "be-persuaded" to "self-persuaded" .

5.2.2 Innovation of the communication process and effect-generated path: from AIDMA to AIIIA

AIDMA was known as traditional advertising model and effect-generated path, in other words, advertising messages aroused attention (Attention), then generated interest to ads (Interest), to generated consumption demand and desire (Desire) and thereby leave a memory in their minds (Memory), until finally purchased the product advertised (Action). AIDMA rules applied to traditional advertisement media with poor interactivity like ratio, television, magazine, newspaper and so on, this passive rules was no longer applicable to interactive advertising.

According to one main conclusion of communication model and effect-generated path, the author pointed AIIIA rules designed for interactive advertising, namely, Attention-Interest-Interaction-Impression-Action. "Interaction", "Impression", "Action" are the key process which is different from traditional AIDMA rules. Firstly, interaction, the core mechanism of interactive advertising, contained different dimensions interaction such as search, browse, click, share and so on. Secondly, impression, which means deeper attitude towards interactive advertising compared to memory, interaction brought not only the memory of advertising messages, but also the in-depth rational and emotional impression in terms of comprehension, brand image, attitude and so on. Finally, the level of action was not only confined to traditional purchasing behaviour, but also sharing interaction in new media era.

5.2.3. New interactive design strategy portfolio

According to the empirical results of this paper, the author provided two strategy combination portfolios for future interactive design practice. One is "High Need for cognition/High Product Involvement-High Human to information interaction", the other is "High Need for Emotion/High Product Involvement-High Human to human interaction". That is to say, advertiser should focus on interaction design, cater to audience with these characteristics ,

getting a positive multiplier effect of advertising effect , for example, advertiser need to design much more human-information interactive options when their target audience are mainly with the characteristics with high need for cognition or high product involvement.

**Authors**

**Ding Wanxing**

Tel. +86 183-1715-9290

Add.  No.516 Jungong Rd.,  Yangpu District, Shanghai,  China. 200093

Master, College of Communication and Art Design, University

 of Shanghai for Science and Technology. The main research

direction is interactive communication and interactive advertising.

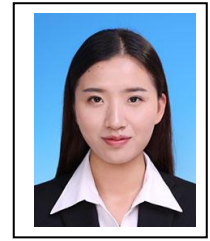